\documentclass[aps,epsfig,manuscript,showpacs]{revtex4}

\usepackage{amsmath,amssymb,graphicx}
\usepackage{epsfig}
\usepackage{graphicx}
\usepackage{enumerate}
\newcommand{\beq}{\begin{equation}}
\newcommand{\eeq}{\end{equation}}

\newcommand{\bea}{\begin{eqnarray}}
\newcommand{\eea}{\end{eqnarray}}

\begin{document}

\title{Numerically exact path integral simulation of nonequilibrium quantum transport and dissipation}

\author{Dvira Segal}
\affiliation{Chemical Physics Theory Group, Department of Chemistry, University
of Toronto, Toronto, Ontario M5S 3H6, Canada}

\author{Andrew J. Millis}
\affiliation{Department of Physics, Columbia University, 538 W 120th St., New York, NY 10027.}

\author{David R. Reichman}
\affiliation{Department of Chemistry, Columbia University, 3000
Broadway, New York, NY 10027}

\begin{abstract}
We develop an iterative, numerically exact approach for the treatment
of nonequilibrium quantum transport and dissipation problems that
avoids the real-time sign problem associated with standard Monte Carlo
techniques. The method requires a well-defined decorrelation time of
the non-local influence functional for proper convergence to the exact
limit. Since finite decorrelation times may arise either from
temperature {\em or from a voltage drop} at zero temperature,
the approach is well suited
for the description of the real-time dynamics of single-molecule
devices and quantum dots driven to a steady-state via interaction with
two or more electron leads. We numerically investigate two
non-trivial models: the evolution of the nonequilibrium population of
a two-level system coupled to two electronic reservoirs, and quantum
transport in the nonequilibrium Anderson model. For the latter case,
two distinct formulations are described.
Results are compared to those obtained by other techniques.
\end{abstract}

 \maketitle

\section{Introduction}

There are several ways in which a quantal entity may exhibit
nontrivial departures from equilibrium. First, a system may evolve
toward equilibrium after application of a transient external pulse or
from a nonequilibrium initial condition. Simple examples of such
situations are now relatively well understood \cite{Legget,Weiss}.
Related to these types of departures from equilibrium, but less well understood,
are more challenging cases of ``quantum quenches,'' whereby the sudden change
of a control parameter induces dynamics that probe non-trivial aspects
of strong correlation or quantum criticality \cite{Bloch}.  
Also underdeveloped is our understanding of quantum mechanical systems driven to nonequilibrium steady-states via coupling to two or more electronic reservoirs.  Since this is the case of direct relevance for the study of transport through quantum dots and molecular electronic devices \cite{QD,MolEl}, the complete description of this type of nonequilibrium behavior is of practical as well as fundamental interest.

There are essentially two main theoretical frameworks for the
calculation of properties related to the approach to, and attainment of, 
nonequilibrium steady-states of the types mentioned above. 
The first is the standard real-time Schwinger-Keldysh technique \cite{Keldysh}.
This approach has led to the exact formulation of steady-state
properties (e.g. the current) in terms of Keldysh Green's functions
\cite{Wingreen}. A variety of direct perturbative and renormalization
group calculations have naturally emerged from this starting point
\cite{Wingreen1,SK1,SK2}.  In addition, real-time Monte Carlo methods
have been formulated on the basis of the Schwinger-Keldysh approach
\cite{MC1,MC2,MC3,Rabani,Marco}.  The Monte Carlo methods are exact in principle
but may be severely limited by numerical sign problems, depending on
the formulation, system and regime under investigation.

The second framework involves the use of Lippmann-Schwinger scattering
states \cite{Lippman} to construct the properties of nonequilibrium
quantum steady-state. This approach has led to several rigorous
results for integrable models \cite{Todorov}.  In the last few years
this viewpoint, combined with the notion of Hershfield's steady-state
density operator \cite{Hershfield}, has inspired the formulation of
new non-perturbative approaches as well as numerical methods
\cite{Scattering,Han,Roy, Hatano}. Most recently, promising numerical
renormalization group approaches have been put forward based directly
on the construction of scattering states \cite{NRG}, and an extension
to the density matrix renormalization group method, incorporating
real-time evolution, has been presented \cite{DMRG1,DMRG2}.

Consideration of more standard classes of nonequilibrium relaxation in
dissipative systems such as the spin-boson model has led to a variety
of path-integral techniques for the numerically exact propagation of
the reduced density matrix of a small system coupled to its
environment \cite{Legget}.  These methods, which include real-time
Monte Carlo techniques \cite{MC1,MC2,MC3} as well as deterministic
iterative approaches \cite{QUAPI,Multiple,Makri-An}, are connected to
the Schwinger-Keldysh type framework discussed above. Here, as in the
Schwinger-Keldysh technique, the approach to equilibrium along a
particular time contour from a prescribed nonequilibrium initial
condition is described. Of these approaches, iterative path-integral
methods have had particular success \cite{QUAPI}. Such methods are
based on the notion that a well defined bath correlation time (if one
exists) renders the range of the influence functional (IF) finite,
allowing for a controlled truncation of memory effects and thus a
deterministic propagation of observables that is free of the real-time
sign problem.

While iterative path-integral approaches have been proven successful
in describing nonequilibrium dynamics in simple spin-boson type models
in the last 15 years, only recently they have been
formulated and used in cases of relevance to transport through quantum
dots and molecular electronic devices \cite{Egger}.  In such systems,
given that a chemical potential difference between electronic
reservoirs leads to a well defined decorrelation time for dynamics
{\em even at zero temperature}, a memory time, beyond which
correlations can be dropped, exists.  This finite-memory
characteristic allows the development of iterative techniques, capable
of describing relaxation in a wide, non-trivial region of parameter
space.

In this work we develop and apply a new iterative path-integral
technique to two models of nonequilibrium transport and dissipation:
the spin-fermion model and the single-impurity Anderson model.  The
techniques developed here hold the potential for the exact description
of long time dynamics in systems driven to a nonequilibrium
steady-state via coupling to two or more electronic reservoirs.  The
method we describe in this work is conceptually similar to the ISPI
approach of Thorwart, Egger and coworkers \cite{Egger}. The distinction
between these two approaches lies mainly in the propagation scheme and the manner in which the
leads are traced out of the problem. In the iterative approach
developed here, the reservoirs are represented as discrete levels and
are eliminated numerically via the Blankenbecler-Scalapino-Sugar (BSS)
identity \cite{BSS}. While this approach has the disadvantage that an 
additional source of systematic error is introduced due to the 
discretization of the lead degrees of freedom, we find empirically 
that the error is easily controlled without undue computational expense.  
The advantage of this approach is that the study of general models 
(for example multi-site Hubbard ``dots'') may be performed with 
essentially no reformulation of methodology. Taking advantage of this fact, we
present a first set of exact results for the out-of-equilibrium
two-lead spin-fermion model. A second difference between the ISPI
approach and the approach outlined in this work is related to the
propagation scheme. Here we combine our matrix formulation with a
propagation scheme similar to that described in \cite{Makri-An}. 
This allows for very efficient propagation that may be trivially 
parallelized with commercially available software \cite{MATLAB}.
These distinctions in scaling and
flexibility of approach render our formulation as a useful compliment
to the previously developed ISPI method.
%

This paper is organized as follows. Section II and Appendix A present
some general aspects of the iterative propagation technique.  Section
III contains a case study of the relaxation of a tunneling system
coupled to two electronic reservoirs. In Section IV we investigate
nonequilibrium transport through an Anderson dot. In Section V we
conclude. We include an alternative formulation of our approach for
the nonequilibrium Anderson dot in Appendix B. This formulation may
also hold promise in related path-integral approaches such as the ISPI
approach. Appendix C describes extensions to finite temperatures.
Finally, Appendix D discusses some aspects of the convergence analysis
which is necessary for elimination of the systematic errors in the
method.

\section{General Formulation of the Iterative Approach}

We consider a generic many-body system, 
consisting of a finite interacting region coupled
to two infinite non-interacting reservoirs.
The Hamiltonian $H$
can be partitioned into a zeroth order term $H_0$ whose solution can
be exactly obtained, typically containing few-body interactions, and
a higher order interaction term $H_1$. We introduce our iterative
approach using the reduced density matrix, $\rho_S={\rm
Tr_B}\{\rho\}$, obtained by tracing the total density matrix $\rho$
over the reservoir degrees of freedom. The time evolution of
$\rho_S(t)$ is exactly given by
\bea
\rho_S(s'',s';t)={\rm Tr_B} \langle s'' | e^{-iHt}\rho(0) e^{iHt} |s' \rangle.
\eea
We decompose the evolution operator into a product of $N$ exponentials,
$e^{iHt}=\left(e^{iH\delta t}\right)^N$; $\delta t=t/N$, and define
the discrete time evolution operator $\mathcal G \equiv e^{iH\delta
t}$. Different Trotter decompositions can be employed for splitting
this operator. For example, we find it convenient to approximate $\mathcal G\sim
e^{iH_1\delta t/2}e^{iH_0\delta t}e^{iH_1\delta t/2}$ when studying
the spin-fermion model (Section III), while for the Anderson model
(Section IV) we find that it is useful to employ a decomposition of the form $\mathcal G\sim e^{iH_0\delta t/2}e^{iH_1\delta t}e^{iH_0\delta
t/2}$. The overall time evolution can be represented
in a path integral formulation,
\bea
&&\rho_S(s'',s',t)=
\int ds_0^+ \int ds_1^+ ... \int ds_{N-1}^+\int ds_0^- \int ds_1^- ... \int ds_{N-1}^-
\nonumber\\
&&{\rm Tr_B}
\Big\{\langle s''|\mathcal G^{\dagger} |s^+_{N-1}\rangle
\langle s_{N-1}^+| \mathcal G^{\dagger}  |s^+_{N-2}\rangle ...
 \langle s_0^+| \rho(0)|s_0^{-}\rangle ...
\langle s^-_{N-2}| \mathcal G  |s^-_{N-1}\rangle
\langle s_{N-1}^-|\mathcal G  |s'\rangle
\Big \},
\label{eq:dynamics}
\eea
where $s_k^{\pm}$ are subsystem (or fictitious) degrees of freedom, representing the
discrete path on the forward ($+$) and backward ($-$) contours.
As an initial condition we may assume that
$\rho(0)=\rho_{B}\rho_S(0)$ with the bath ($B$) uncoupled to the subsystem.
In what follows we refer to the integrand in (\ref{eq:dynamics}) as an "Influence Functional" (IF) \cite{FV},
and denote it by $I(s_0^{\pm},s_1^{\pm}...s_{N}^{\pm})$, assigning $s_N^+=s''$, $s_N^-=s'$. Note that our definition of the IF is more general than that contained in the original work of Feynman and Vernon \cite{FV}. 
We chose this loose definition to make connection with the iterative schemes developed in the previous path-integral based numerical work \cite{QUAPI}.

The IF combines the information of subsystem and bath degrees of
freedom with system-bath interactions, and its form is analytically
known only in special cases. For example, for a harmonic bath
bilinearly coupled to a subsystem the IF is an exponential of a
quadratic form, multiplied by free subsystem propagation terms
\cite{FV}
\bea
I^{har}(s_0^{\pm}...s_{N}^{\pm})&=&\exp\Big[-\sum_{k}^N \sum_{k'=0}^k(s_k^+-s_k^-)(\eta_{k,k'}s_{k'}^+ -   \eta_{k,k'}^*s_{k'}^- ) \Big]
\nonumber\\
&\times&
\langle s_N^+|e^{-iH_0\delta t} |s_{N-1}^+\rangle ... \langle s_0^+ |\rho_S(0)|s_0^-\rangle ...
\langle s_{N-1}^-| e^{iH_0\delta t} | s_N^-\rangle.
\label{eq:harmonic}
\eea
The coefficients $\eta_{k,k'}$ depend on the bath spectral function
and the temperature \cite{QUAPI}. For a general anharmonic environment
the IF may contain multiple-site interactions, where the
coefficients are not known in general \cite{Multiple}. However, even
when the form of the IF is analytically known as in
(\ref{eq:harmonic}), it still combines long range interactions
limiting brute force direct numerical simulations to very short times.

For a system coupled to a {\it single} thermal reservoir this
challenge has been tackled at finite temperatures where a natural bath decoherence time exists. As noted by Makri and Makarov \cite{QUAPI}, such cases are characterized by the useful feature that nonlocal correlations contained in the IF decay exponentially, enabling a
(controlled) truncation of the IF that includes only a finite memory
length. Based on this feature, an iterative scheme for evaluating
the (finite dimensional) path integral has been developed \cite{QUAPI}.
While the original quasi-adiabatic path integral (QUAPI) algorithm
was developed based on the analytical pairwise form of the IF
specific to harmonic reservoirs (\ref{eq:harmonic}), a subsequent more general approach proposed in Ref. \cite{Makri-An} is based only on the fact that memory effects at finite temperatures generically vanish exponentially in the long time limit.

This idea can be further employed to simulate the dynamics of a
generic {\it nonequilibrium} bias-driven system \cite{Egger}. Since
in standard nonequilibrium situations bath correlations die
exponentially, the IF can be truncated
beyond a memory time $\tau_c=N_s \delta t$, corresponding to the
time where beyond which bath correlations may be controllably ignored. Here, $N_s$ is an integer, $\delta t$ is the discretized time step, and $\tau_c$ is a correlation time dictated by the nonequilibrium situation. 
For a system under a dc potential bias $\Delta\mu$ at zero temperature, $\tau_c\sim 1/\Delta\mu$, while at temperatures for which $T > \Delta\mu$ temperature sets the scale of the memory range. We therefore 
write the total influence functional approximately as
\bea I(s_0^{\pm},s_1^{\pm},s_2^{\pm},...,s_N^{\pm})\approx
 I(s_0^{\pm},s_1^{\pm},...,s_{N_s}^{\pm})
I_s(s_1^{\pm},s_2^{\pm},...,s_{N_s+1}^{\pm}) ...
I_s(s_{N-Ns}^{\pm},s_{N-N_s+1}^{\pm},...,s_N^{\pm}). \label{eq:IF}
\eea
with
\bea I_s(s_k,s_{k+1},...,s_{k+N_s})=
\frac{I(s_k^{\pm},s_{k+1}^{\pm},...,s_{k+N_s}^{\pm})}{I(s_{k}^{\pm},s_{k+1}^{\pm},...,s_{k+N_s-1}^{\pm})}.
\label{eq:Is} \eea
%
The errors in Eq. (\ref{eq:IF}) are the usual Trotter error arising from the time 
discretization and the truncation to a finite memory time 
$\tau_c=N_s \delta t$. Both of these errors can be controlled.
%
Eq. (\ref{eq:IF}) can be understood as a simple generalization of the pairwise
expression (\ref{eq:harmonic}) for which
\bea I_s^{har}(s_k^{\pm},s_{k+1}^{\pm},..,s_{k+N_s}^{\pm}) =
f_0(s_k^{\pm}) f_1(s_k^{\pm},s_{k+1}^{\pm})...
f_{N_s}(s_k^{\pm},s_{k+N_s}^{\pm}). \eea
The one-body and two-body functions $f$ can be obtained by
rearranging Eq. (\ref{eq:harmonic}). From these expressions we
recursively build the finite-range IF for a general model. We assume
that the complete functional decays to zero with time constant $\tau_c=N_s\delta t$,
($N_s<N$), thus it can be approximated by the product
\bea &&I(s_0^{\pm},s_1^{\pm},s_2^{\pm},...,s_N^{\pm})\approx
\nonumber\\
&& I(s_0^{\pm},s_1^{\pm},s_2^{\pm},...,s_{N-1}^{\pm}) \frac
{I(s_1^{\pm},s_2^{\pm},s_3^{\pm},...,s_{N}^{\pm}) }
{I(s_1^{\pm},s_2^{\pm},s_3^{\pm},...,s_{N-1}^{\pm}) }.
\label{eq:decomp} \eea
%
By recursively applying this rule, the truncated IF is further decomposed until it correlates
interactions within $\tau_c$ only,
\bea
&&I(s_0^{\pm},s_1^{\pm},s_2^{\pm},...,s_N^{\pm})\approx \nonumber\\
&& I(s_0^{\pm},s_1^{\pm},...,s_{N_s}^{\pm})
\frac{I(s_1^{\pm},s_2^{\pm},...,s_{N_s+1}^{\pm})}
{I(s_1^{\pm},s_2^{\pm},...,s_{N_s}^{\pm})}
\frac{I(s_2^{\pm},s_3^{\pm},...,s_{N_s+2}^{\pm})}
{I(s_2^{\pm},s_3^{\pm},...,s_{N_s+1}^{\pm})}...
\frac{I(s_{N-N_s}^{\pm},s_{N-N_s+1}^{\pm},...,s_{N}^{\pm})}
{I(s_{N-N_s}^{\pm},s_{N-N_s+1}^{\pm},...,s_{N-1}^{\pm})},
\label{eq:IFbreak}
 \eea
resulting in Eqs. (\ref{eq:IF}) and (\ref{eq:Is}).
The physical content of this approach, which is similar to that described in \cite{Makri-An}, 
is outlined in Appendix A.
The approach becomes exact as $\tau_c\rightarrow \infty$.
Outside of the initial propagation step, $I_s(s_0^{\pm},s_1^{\pm},...,
s_{N_s}^{\pm})\equiv I(s_0^{\pm},s_1^{\pm},...,s_{N_s}^{\pm})$, we
can identify the functions $I_s$ [Eq. (\ref{eq:IF})] as the ratio
between two IFs where the numerator is calculated with an additional
time step, Eq. (\ref{eq:Is}).
Next, based on the decomposition (\ref{eq:IF}) we can iteratively integrate Eq. (\ref{eq:dynamics})
by defining a multiple-time reduced density matrix $\tilde \rho_S(s_{k},s_{k+1},..,s_{k+N_s-1})$.
Its initial value is given by $\tilde \rho_S(s_0^{\pm},... ,s_{N_s-1}^{\pm})=I$,
and it is evolution is dictated by
\bea
\tilde \rho_S(s_1^{\pm},...,s_{N_s}^{\pm})=
\int ds_0^{\pm} \tilde\rho_S(s_0^{\pm},...,s_{N_s-1}^{\pm})I_s(s_0^{\pm},...,s_{N_s}^{\pm}),
\label{eq:prop0}
\eea
with
\bea
I_s(s_0^{\pm},...,s_{N_s}^{\pm})=  {\rm Tr_B} \{
\langle s_{N_s}^+|\mathcal G^{\dagger} | s_{N_s-1}^+\rangle ...
\langle s_1^+|\mathcal G^{\dagger} | s_0^+\rangle
\langle s_0^+ | \rho(0) |s_0^- \rangle
\langle s_0^-|\mathcal G | s_1^-\rangle...
\langle s_{N_s-1}^-|\mathcal G | s_{N_s}^-\rangle \}.
\nonumber\\
\label{eq:prop1}
\eea
A general propagation step involves integration over two ($\pm$) coordinates,
\bea
\tilde \rho_S(s_{k+1}^{\pm},...,s_{k+N_s}^{\pm})=
\int ds_{k}^{\pm} \tilde\rho_S(s_{k}^{\pm},...,s_{k+N_s-1}^{\pm})
I_s(s_{k}^{\pm},...,s_{k+N_s}^{\pm}),
\label{eq:prop2}
\eea
where the time-local ($t_k=k\delta t$) reduced density matrix is obtained by summing over all intermediate states,
\bea
\rho_S(t_k)=\int ds_{k-1}^{\pm} ...ds_{k-N_s+1}^{\pm} \tilde \rho_S(s_{k-N_s+1}^{\pm},...,s_{k}^{\pm}).
\label{eq:prop3}
\eea
%
The evolution at shorter times $k<N_s$ can be calculated in a numerically exact way. 
Before turning to specific models we would like to make the following comments regarding the above derivation. (i) The specific partitioning of the Hamiltonian into $H_0$ and $H_1$ depends on the model investigated. As we show below, $H_0$ may include only the
subsystem degrees of freedom (spin-fermion model), or it may be
constructed involving all two-body terms (Anderson model). (ii)
Obviously, the decomposition (\ref{eq:decomp}) is not unique,
however, different schemes should lead to equivalent time evolution, 
and thus the partitioning is a matter of numerical convenience.
(iii) The truncated IF ($I_s$) is not necessarily a time invariant. As
we show below, in the spin-fermion model $I_s$ does not depend on
time, thus in this case it needs to be evaluated only once during the propagation scheme. 
In contrast for the Anderson model standard use of the Hubbard-Stratonovich transformation leads to an IF expression that has to 
be updated at each time step. 
In Appendix C we outline an approach that does not make use of the Hubbard-Stratonovich 
transformation and thus produces a form on the IF of the Anderson model that is time-independent. 
(iv) The short-range function $I_s$ can be analytically evaluated in some special cases \cite{QUAPI,Multiple}. 
For general reservoirs it may be evaluated numerically, by using finite size reservoirs 
as described in the next section. (v) The approach outlined here is not 
restricted to specific statistics of the leads (boson or fermion) and is solely based on the fact that at
finite temperature and/or finite bias bath correlations
exponentially decay at long time. 
Therefore, it can be used to treat finite temperature anharmonic 
bosonic environments \cite{Makri-An} as well as nonequilibrium Fermi systems.

\section{Dissipation in the Nonequilibrium Spin-Fermion Model}

\subsection{Model}

As a first example, we consider the dynamics of a two-state system
coupled to two fermionic leads maintained at different chemical
potential values, the "spin-fermion model" (SF). This model has been
considered in a series of recent papers
\cite{Mitra-spin,Marcus,MitraC,DasSarma}, and serves as a simple,
albeit non-trivial, example exhibiting the generic behavior
associated with the approach to a nonequilibrium steady-state. In
particular, at zero temperature the chemical potential difference
$\Delta \mu$ sets the essential energy scale for dephasing as is
expected generically in more complex models such as the
nonequilibrium Kondo model \cite{KondoNoneq}. It should be noted,
however, the connection between the model studied here and the
nonequilibrium Kondo model \cite{KondoNoneq} is more tenuous then
that between the tunneling center model in equilibrium \cite{Legget} 
and the standard (equilibrium) Kondo model \cite{Kondo}. We take as our
Hamiltonian
\bea
H^{SF}&=&H_0+H_1;
\nonumber\\
H_0&=&H_S;  \,\,\,\,\,\, H_1=H_B+H_{SB}.
\eea
The bath Hamiltonian $H_{B}$ is taken to be that of two independent
leads ($\alpha$=$L,R$) characterized by (spinless) free-fermion
statistics with different chemical potentials, namely
\begin{equation}
H_B=\sum_{\alpha,k}\epsilon_k c_{\alpha,k}^{\dagger} c_{\alpha,k}.
\label{eq:HB}
\end{equation}
The operator $c_{\alpha,k}^{\dagger}$ ($c_{\alpha,k}$) creates (annihilates)
an electron with momentum $k$ in the $\alpha$-th lead. The system
Hamiltonian $H_S$ consists of a two-level system (TLS) with a bare
tunneling amplitude $\Delta$ and a level splitting $B$,
\begin{equation}
H_S=\frac{B}{2}\sigma_z+\frac{\Delta}{2} \sigma_x.
\end{equation}
We take the general form for the system-bath coupling to be
\begin{equation}
H_{SB}=\sum_{\alpha,\alpha',k,k'}V_{\alpha,k;\alpha',k'}c_{\alpha,k}^{\dagger}c_{\alpha',k'}\sigma_z.
\end{equation}
Different versions of the model may be expressed via different forms
of the coupling parameters $V$. In this paper we focus on the model
presented in Ref. \cite{Ng,Marcus}, where the momentum dependence of the scattering potential is neglected. The system-bath scattering potentials are then given by $V_{\alpha,\alpha'}$, where $\alpha,\alpha'=L,R$ are the Fermi sea indices.

In the standard application of iterative path-integral approaches,
two features greatly simplify the propagation algorithm. First, the
form of the Feynman-Vernon influence functional is known
analytically. Second, the influence functional is pair-wise
decomposable \cite{QUAPI}. As discussed in the previous section, 
neither of these features is necessary for the numerical implementation of an efficient iterative routine.

Recently, the analytical structure of the influence functional in the 
spin-fermion model considered here has been elucidated, with a
modified pair-wise Coulomb gas behavior emerging at long times \cite{MitraC}. 
However, our recent numerical results have illustrated that in some cases for 
strong coupling of the system to the leads, most of the relevant dynamical 
evolution occurs in time intervals {\em before} strict Coulomb gas behavior holds \cite{Marcus}.

The exact dynamics follows Eq. (\ref{eq:dynamics}).
Assuming separable initial conditions $\rho(t=0)=\rho_S(t=0)\rho_B(t=0)$,
we can identify the IF in the present model as
\bea &&I(s_0^{\pm}, s_1^{\pm}, ... ,s_N^{\pm})=
 \langle s_0^+ | \rho_S(0)|s_0^- \rangle
K(s_N^{\pm},s_{N-1}^{\pm})... K(s_2^{\pm},s_1^{\pm})K(s_1^{\pm},s_0^{\pm})  \times
\nonumber\\
&&{\rm Tr}_B\Big\{
e^{-iH_1(s_N^+) \delta t/2} e^{-iH_1(s_{N-1}^+) \delta t}...
 e^{-iH_1(s_{0}^+) \delta t/2} \rho_B(0) e^{iH_1(s_0^-) \delta
t/2}....
e^{iH_1(s_{N-1}^-) \delta t} e^{iH_1(s_{N}^-) \delta t/2} \Big\}.
\label{eq:IFSF}
\eea
where  $H_1=H_B+H_{SB}$ provides an adiabatic partitioning of the
Hamiltonian, $s_k^{\pm}$ are forward ($+$) and backward ($-$) spin states along the paths, and $K(s_{k+1}^{\pm},s_{k}^{\pm}) =\langle s_{k+1}^+|e^{-iH_S \delta t}| s_{k}^+\rangle \langle s_{k}^- |e^{iH_S\delta t}| s_{k+1}^-\rangle$ is the propagator matrix for the isolated subsystem.

The reduced density matrix is time-propagated by employing the
iterative scheme (\ref{eq:prop0})-(\ref{eq:prop3}), where the
function $I_s$ [Eq. (\ref{eq:Is})] is calculated by taking ratios of
the corresponding truncated IF (\ref{eq:IFSF}). Note that this
function  is time-translationally invariant, thus we need to calculate it only once.

\subsection{Results}

To numerically calculate the influence functional, we express the
lead Hamiltonians in terms of a finite number of fermions. Then, as
in the standard BSS Monte Carlo approach to lattice fermions \cite{BSS}, 
the resulting trace may be expressed as a simple determinant containing the 
1-body matrices that represent exponentials of operators that are quadratic 
in fermionic creation and annihilation operators. 
It should be noted that this discretization of the bath leads to systematic 
error in the results, unlike the case for the related ISPI
approach of Thorwart, Egger and coworkers \cite{Egger}. 
However, the discretized approach for tracing out the bath is more flexible in that 
cases where the analytic structure of the self-energy terms, such as structured "dot" 
with several correlated sites, may be easily treated. 
Furthermore, bosonic analogs of generalized Anderson models may be treated easily as well \cite{BA}, 
using the boson version of the BSS formalism \cite{Klich}. 
This fact may be of importance for the recently developed bosonic versions of DMFT \cite{BV,Anders}, 
where for out-of-equilibrium situations or at finite temperatures the approach 
outlined here may potentially serve as a real-time impurity solver. 
Fortunately, since the time intervals over which the bath is "measured" are short, 
we have found that the infinite bath result is easy reached even with a relatively small number of 
effective bath fermions $\sim 40$.

We use the following parameters: $\Delta=1$, $B=0$,
$\Delta\mu\sim0.5-2$, and $\rho
V_{\alpha,\alpha'}=\lambda(1-\delta_{\alpha,\alpha'})$, considering
only inter-bath system-bath couplings, where spin
polarization is coupled to scattering events between the
nonequilibrium reservoirs. For simplicity we assume zero
temperature. The generalization to finite temperature is
straightforward as outlined in Appendix B. Since the iterative approach outlined above requires a finite range of memory for the influence functional, we work with a bias large enough 
to ensure facile convergence in the numerical examples outlined below.

\begin{figure}[htbp]
\vspace{0mm} \hspace{0mm}
{\hbox{\epsfxsize=90mm\epsffile{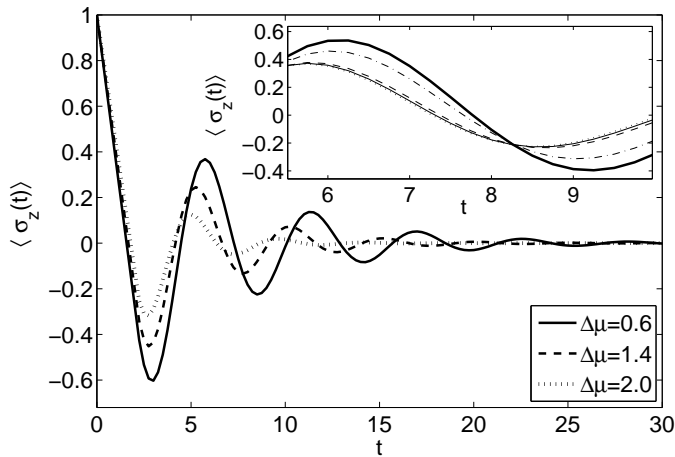}}} \caption{Polarization in the nonequilibrium
spin-fermion model at different values of the bias voltage
$\Delta\mu$=0.6 (full); $\Delta\mu$=1.4 (dashed); $\Delta\mu$=2
(dashed-dotted), $B=0$, $\Delta$=1, $\lambda$=0.2, $\delta t=0.25$,
$N_s=8$. Inset:  convergence with increasing correlation
time  at $\Delta \mu=0.6$, $N_s=3$ (dark full); $N_s=4$ (dashed-dotted);
$N_s=7$ (dashed); $N_s=8$ (dotted); $N_s=9$ (light full).
Data was generated using 80 states per bath, which is sufficient to ensure convergence in the regime of parameters presented here.}
\label{FigSF1}
\end{figure}

In Fig. \ref{FigSF1} we show the dynamics of the spin polarization
$\langle \sigma_{z}(t) \rangle$ for several different values of the
bias $\Delta \mu$, distributed symmetrically between the $L$ and $R$ leads. The role of the chemical potential difference as
a temperature-like contributor to dephasing is clear \cite{Marcus}.
We analyze (inset) the memory error in our algorithm by increasing
$\tau_c$, keeping $\delta t$ fixed. As expected, we find that
$\tau_c$ roughly corresponds to $1/\Delta \mu$. Thus, for
$\Delta\mu\sim 1$, taking $\delta t=0.25$, the dynamics is
converging for $N_s\gtrsim5$.  A complete discussion of the appropriate
convergence analysis is presented in Appendix D for the Anderson model.

\begin{figure}[htbp]
\vspace{0mm} \hspace{0mm} {\hbox{\epsfxsize=90mm \epsffile{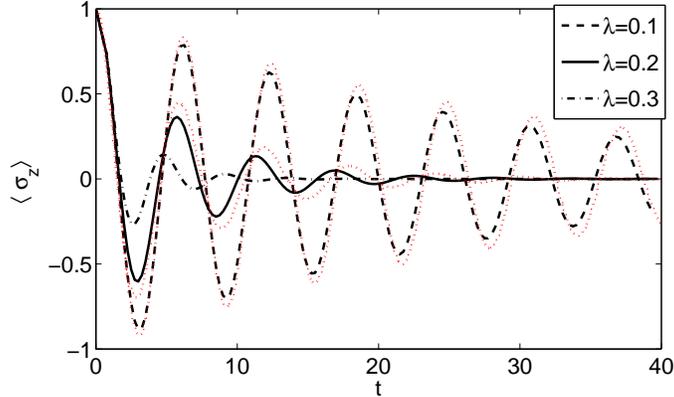}}}
\caption{Polarization in the nonequilibrium spin-fermion model at
different spin-bath couplings, $\lambda$=0.1 (dashed); $\lambda=0.2$
(full); $\lambda=0.3$ (dotted). 
Here $\Delta\mu=0.6$ and $\delta t=0.25$, $N_s=10$. The dotted line was generated using a  
nonequilibrium version of the "non-interacting spin-blip approximation" 
\cite{Mitra-spin,Marcus}. } \label{FigSF2}
\end{figure}


In Fig. \ref{FigSF2} we compare our numerically exact results, with
the results of a generalized "non-interacting blip" approximation as
formulated by Mitra and Millis \cite{Mitra-spin,Marcus}. While at
weak coupling the dynamics reasonably agree, for strong
interactions $\lambda=0.3$ ($\pi \rho V_{\alpha,\alpha'} \sim 1$) the perturbative method diverges \cite{Marcus}.  We found that at weak to intermediate interaction strengths our results systematically converge with increasing memory time $\tau_c$.
For strong interactions $\pi \delta \sim 1$ the time-step in our simulations should be made further smaller $\delta
t\sim0.1$ in order to achieve convergence, demanding extensive
computation effort as $N_s>16$ for $\Delta\mu\sim0.6$.

It would be most useful to undertake a systematic study of the
dynamical phase diagram in $(T,\Delta \mu, \rho V_{\alpha,\alpha'})$
space in the regions where our iterative technique is convergent.
Such a study would be quite useful for the understanding of the
approach to nonequilibrium steady state, and will be the subject of
a future investigation.

\section{Nonequilibrium Transport Through an Anderson Dot}

\subsection{Method}

The single impurity Anderson Model (SIAM)  \cite{Anderson}
is one of the most important and well-studied models in condensed matter physics. While it was originally introduced to describe the behavior of magnetic impurities in non-magnetic hosts \cite{Kondo},
it has more recently served as a general model for understanding
transport in correlated nanoscale systems \cite{QD,MolEl}.  In such cases, the impurity is hybridized with more than one reservoir, and if the chemical potentials of the reservoirs are not identical, nonequilibrium transport will occur.  Here, we present a numerically exact scheme for calculating dynamical quantities such as the time-dependent occupation and current in such systems.
The approach outlined in this section relies on the discrete Hubbard-Stratonovich transformation. An alternative and more general approach is outlined in Appendix C that does not employ this transformation. While the approach of Appendix C offers several advantages, it is somewhat simpler to implement the scheme described here, and for that reason we follow it for the sake of illustrative calculation.

The SIAM  model includes a resonant level of energy $\epsilon_d$, described by the creation operator $d^{\dagger}_{\sigma}$
($\sigma=\uparrow,\downarrow$ denotes the spin orientation)
coupled to two fermionic leads ($\alpha=L,R$) of different chemical
potentials $\mu_\alpha$,
\bea
H^{AM}&=&\sum_{\sigma} \epsilon_d d_{\sigma}^{\dagger}d_{\sigma} +
U d^{\dagger}_{\uparrow}d_{\uparrow} d^{\dagger}_{\downarrow}d_{\downarrow}
\nonumber\\
&+&\sum_{\alpha,k,\sigma}\epsilon_k
c_{\alpha,k,\sigma}^{\dagger}c_{\alpha,k,\sigma}+\sum_{\alpha,k,\sigma}V_{\alpha,k}c_{\alpha,k,\sigma}^{\dagger}d_{\sigma}
+h.c.
\label{eq:SIAM}
\eea
Here $c_{\alpha,k,\sigma}^{\dagger}$ ($c_{\alpha,k,\sigma}$) denotes the creation (annihilation) of an
electron with momentum $k$ and spin $\sigma$ in the $\alpha$ lead,
$U$ stands for the onsite repulsion energy, and
$V_{\alpha,k}$ are the impurity-$\alpha$ lead coupling elements.
The Hamiltonian (\ref{eq:SIAM}) can be also rewritten as $H^{AM}=H_0+H_1$, where
$H_0$ includes the exactly solvable non-interacting part, and $H_1$ includes the many-body term,
\bea
H_0&=& \sum_{\sigma} (U/2+\epsilon_d)
d_{\sigma}^{\dagger}d_{\sigma}
 \nonumber\\
&+&\sum_{\alpha,k,\sigma}\epsilon_k
c_{\alpha,k,\sigma}^{\dagger}c_{\alpha,k,\sigma}+\sum_{\alpha,k,\sigma}V_{\alpha,k}c_{\alpha,k,\sigma}^{\dagger}d_{\sigma}
+h.c.
\nonumber\\
 H_1&=& U\big[n_{d,\uparrow}n_{d,\downarrow} -\frac{1}{2} (n_{d,
\uparrow} +n_{d,\downarrow})\big]. \label{eq:H} \eea
Here $n_{d,\sigma}=d^{\dagger}_{\sigma}d_{\sigma}$ is the impurity
occupation number operator. The shifted single-particle energies are
denoted by $E_d=\epsilon_d+U/2$. We also define $\Gamma=\sum_{\alpha}\Gamma_{\alpha}$, where $\Gamma_{\alpha}=\pi
\sum_{k}|V_{\alpha,k}|^2\delta(\epsilon-\epsilon_k)$ is the
hybridization energy of the resonant level with the $\alpha$ metal.

Our objective is to calculate the dynamics of a quadratic
operator $\hat A$, either given by system or bath degrees of
freedom. This can be generally done by studying the Heisenberg
equation of motion of the exponential operator $e^{\lambda \hat A}$ 
with $\lambda$ here a variable that is taken to vanish at the end of the calculation,
\bea
&&\langle \hat A(t) \rangle = {\rm Tr} (\rho \hat A) = \nonumber\\ &&
\lim_{\lambda \rightarrow 0} \frac{\partial}{\partial \lambda} {\rm
Tr}\big[\rho(0) e^{iH^{AM}t}e^{\lambda \hat A}e^{-iH^{AM}t}  \big].
\label{eq:At}
\eea
Here $\rho$ is the total density matrix. For simplicity, we assume
that at the initial time ($t=0$) the dot and the bath are decoupled,
the impurity site is empty, and the bath is prepared in a
nonequilibrium (biased) zero temperature state. The time evolution
of $\hat A$ can be obtained following a scheme analogous to that
outlined in Section II for the reduced density matrix. For clarity,
we re-derive an explicit expression for the generalized IF in the present case as well.

First we use a standard factorization of the time evolution operator
$e^{iH^{AM}t} = (e^{iH^{AM}\delta t})^N$,
and assume the Trotter decomposition $e^{iH_{AM}\delta t}\approx\big(
e^{iH_0\delta t/2} e^{iH_1 \delta t} e^{iH_0\delta t/2}  \big)$.
The many body term $H_1$ is further eliminated by introducing auxiliary Ising variables $s=\pm$ via the Hubbard-Stratonovich transformation \cite{Hubb-Strat},
\bea
e^{iH_1 \delta t} &=& \frac{1}{2} \sum_{s}
e^{-s \kappa_+ (n_{d,\uparrow}-n_{d,\downarrow}) }; \nonumber\\
e^{-iH_1 \delta t} &=& \frac{1}{2} \sum_{s} e^{-s
\kappa_-(n_{d,\uparrow}-n_{d,\downarrow})}, \eea
where $\kappa_{\pm}=\kappa' \mp i \kappa'' $,
$\kappa'=\sinh^{-1}[\sin(\delta t U/2)]^{1/2}$, $\kappa''
=\sin^{-1}[\sin (\delta t U/2)]^{1/2}$.
The uniqueness of this transformation requires $U \delta t < \pi$.
In what follows we use the following notation
\bea e^{H_\pm(s)} &\equiv&
e^{-s\kappa_\pm(n_{d,\uparrow}-n_{d,\downarrow})}. \label{eq:Hpm}
\eea
Incorporating the Trotter decomposition and the HF transformation (\ref{eq:Hpm}) into Eq. (\ref{eq:At}), we find that at zero temperature the time evolution of $\hat A$ is given by
\bea
&&\langle \hat A(t)\rangle =
\nonumber\\
&&\lim_{\lambda \rightarrow 0} \frac{\partial}{\partial \lambda}
\big \langle 0\big| \left( e^{iH_0\delta t/2} e^{iH_1 \delta t}
e^{iH_0\delta t/2}  \right)^N e^{\lambda \hat A}
 \left( e^{-iH_0\delta t/2} e^{-iH_1 \delta t}  e^{-iH_0\delta t/2}   \right)^N
  \big |0 \big \rangle
\nonumber\\
&=&
\lim_{\lambda \rightarrow 0} \frac{\partial}{\partial \lambda} \Big \{
\frac{1}{2^{2N}}\int ds_1^{\pm} ds_2^{\pm}... ds_N^{\pm} \big \langle
0\big | \left( e^{iH_0/2 \delta t} e^{H_+(s_N^+)}    e^{iH_0/2 \delta t}  \right)
 ...  \left( e^{iH_0 \delta t/2} e^{H_+(s_1^+)}    e^{iH_0 \delta t/2}
\right)
\nonumber\\
&\times& e^{\lambda \hat A} \times \left(  e^{-iH_0 \delta t/2}  e^{H_-(s_1^-)} e^{-iH_0 \delta t/2}
\right) ...   \left( e^{-iH_0 \delta t/2}   e^{H_-(s_N^-)} e^{-iH_0 \delta t/2} \right)
 \big| 0 \big \rangle \Big \},
 \label{eq:At2}
\eea
where $|0 \rangle$ is the initial (zero temperature) state of the
total system. For convenience, we evaluate Eq. (\ref{eq:At2}) by
diagonalizing the Hamiltonian $H_0$ [see Eq. (\ref{eq:H})], and
rewriting $H_{\pm}$ in terms of the new basis
\bea \bar H_0&=& \sum_{\nu}\epsilon_{\nu} b_{\nu}^{\dagger}b_{\nu} ;
\,\,\,
H_0=V\bar H_0 V^{-1} \nonumber\\
\bar H_{\pm}&=&\sum_{\nu,\nu'}
\beta_{\nu}^*\beta_{\nu'}b_{\nu}^{\dagger}b_{\nu'}, \eea
with $\beta_{\nu}$ as the transformation matrix elements. We further transform both the operator of interest and the ground state
into the new representation $ {\hat A}= V \bar {\hat A} V^{-1}$,
 $|\bar 0\rangle = V^{-1}|0\rangle$.
The IF is identified as the integrand in (\ref{eq:At2}), where we
truncate interactions beyond the memory time $\tau_c=N_s\delta t$,
\bea
&& I(s_k^{\pm},... s_{k+N_s}^{\pm})=
\nonumber\\
&&\frac{1}{2^{2 N_s}}\big \langle \bar 0\big| \mathcal
G_+(s_{k+N_s}^+) ... \mathcal G_+(s_{k}^+) e^{i\bar H_0 (k-1) \delta
t} e^{\lambda \bar {\hat A}} e^{-i\bar H_0 (k-1) \delta t} \mathcal
G_-(s_{k}^-) \mathcal G_-(s_{k+N_s}^-) \big |\bar 0\big \rangle,
\nonumber\\
\label{eq:IFAM}
\eea
with $\mathcal G_{+}(s_k^{\pm}) = \left( e^{i\bar H_0 \delta t/2}
e^{\bar H_{+}(s_{k}^{\pm})}  e^{i\bar H_0 \delta t/2} \right)$
and $\mathcal G_-=\mathcal G_+^{\dagger}$.
Finally, we can build the function $I_s$  [Eq. (\ref{eq:IF})] using (\ref{eq:Is}), and the operator of interest ${\hat A}$ may be propagated using a scheme analogous to that developed for the reduced density matrix, Eqs. (\ref{eq:prop0})-(\ref{eq:prop3}).

Before presenting numerical results we make the following
comments. First, in the present scheme the IF needs to be updated at
each time step since the truncated IF [Eq. (\ref{eq:IFAM})] explicitly
depends on the present time $t_k=k\delta t$. Second, the operator
$\hat A$ can represent various quadratic operators.  Thus quantities such as the impurity population or the current through the junction \cite{Rabani} may be investigated on the same footing.

\subsection{Results}

The IF (\ref{eq:IFAM}) is the core of our calculation. It is
evaluated numerically using the zero temperature relationship
$\langle 0| e^B |0 \rangle = \det[e^b]_{occ.}$, where $b$ is a
single particle operator, $B=\sum b$, and the determinant is carried
over occupied states only. Extensions to finite temperature are
standard, see Appendix B. Similarly to the spin-fermion model we represent the reservoirs by a finite set of fermions, with energies determined by the metals' dispersion relation.  Calculations must be converged with respect to the number of discrete lead states.
The $\lambda$ derivative in (\ref{eq:At2}) is handled numerically, by calculating the IF for several (small) values of $\lambda$.

In the following we typically use the following conventions and parameters: a symmetrically distributed voltage bias between two leads with $\Delta\mu=0.4-0.6$, a reservoir bandwidth of $D=1$, a resonant level energy $E_d=0.3$, and hybridization strength $\Gamma_{\alpha}$=0.025-0.1. 
Note that the actual hybridization parameter utilized in the simulations is the coupling $V_{\alpha,k}=\sqrt{\Gamma_{\alpha}/\pi\rho_{\alpha}}$, where $\rho_{\alpha}$ is the density of states of the $\alpha$ lead. For these parameters we find that convergence is achieved using $L\leq 240$ states per spin per bath. We have also verified that for $\Delta\mu=0.4$ the memory time $\tau_c\sim 3.2$ leads to convergence with $\delta t=0.8$ and $N_s=4$, provided $\frac{U}{\Gamma} \lesssim 3$ (see Appendix D).

\begin{figure}[htbp]
\vspace{0mm} \hspace{0mm} {\hbox{\epsfxsize=90mm
\epsffile{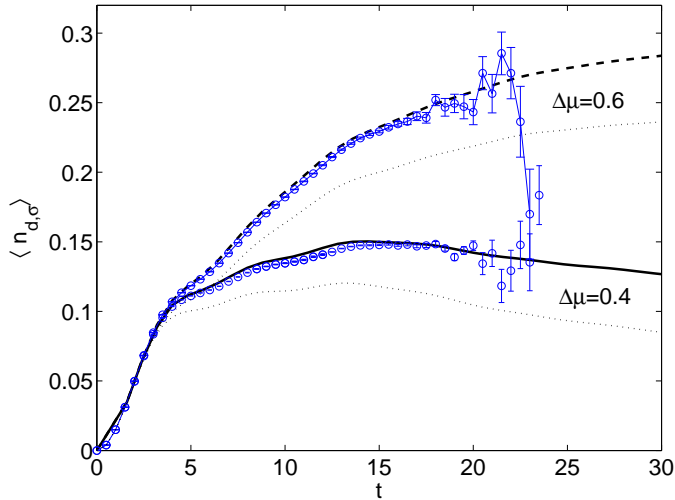}}} \caption{Resonant level dynamics at different
values of the voltage bias, $\Delta\mu=0.6$ (dashed);
$\Delta\mu=0.4$ (full). $U=0.1$, $\Gamma_{\alpha}$=0.025, $E_d=0.3$,
$\tau_c=3.2$. The dotted lines show for reference
the exact $U=0$ dynamics at $\Delta\mu=0.6, 0.4$, (top to bottom).
The circles are the respective Monte Carlo points. Calculations are performed at $T=0$, while Monte Carlo data utilizes $T=1/200$ which is effectively converged to the $T=0$ limit.} \label{Fig0}
\end{figure}


We begin by investigating the dynamics for a relatively small
interaction $U=0.1$ ($\Gamma\equiv \Gamma_L+\Gamma_R$ and $U/\Gamma=2$). 
In this regime we are able to systematically converge the results of our 
procedure with respect to the three sources of systematic error, 
namely those associated with time step and bath discretization as well 
as non-local memory truncation. 
Figure \ref{Fig0} presents the time evolution of the dot occupation for two different bias voltages, 
$\Delta\mu=0.6$ (dashed) and $\Delta\mu=0.4$ (full), assuming the dot ($E_d=0.3$) is initially empty. 
The results are compared to exact real-time Monte Carlo (MC) simulations employing 
the hybridization expansion \cite{Phillip} manifesting good agreement at this relatively small $U$: At short times the IF data reproduce the MC features, while close to steady-state the MC results become
increasingly unstable. 
The more recently developed weak-coupling expansion \cite{Phillip2} 
is capable of significantly extending the time regime for which converged results may be 
obtained via Monte Carlo for symmetric cases, however this restriction limits 
the cases for which long-time results may be obtained. 
The MC data presented in this paper was generated at finite-low temperature,  $1/T=200$.
We have verified (data not included) that for this temperature range the population dynamics essentially 
coincide with the strictly zero temperature case. 
The extremely small deviations between MC data and our approach at $U=0.1$ in Fig. \ref{Fig0} 
are the result of small differences in temperature and the fact that a sharp, finite band 
is assumed in our calculations.

\begin{figure}[htbp]
\vspace{0mm} \hspace{0mm}
 {\hbox{\epsfxsize=90mm \epsffile{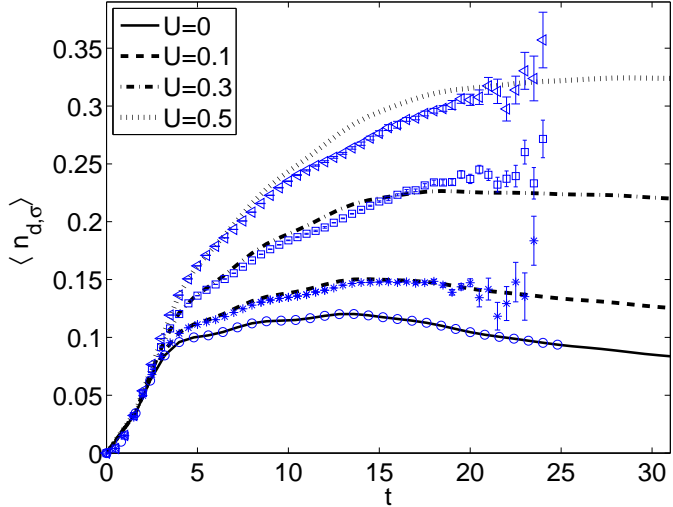}}}
  \caption{
Population of the resonant level in the  Anderson model.
The results for $U=0$ (full), $U=0.1$  (dashed), $U=0.3$ (dashed-dotted), $U=0.5$ (dotted) are compared with the exact dynamics at $U=0$ ($\circ$) and Monte Carlo data (*, $\square$, $\triangleleft$). The physical parameters of the model are $D=1$, $\Delta \mu=0.4$, $E_d=0.3$ and $\Gamma_{\alpha}$=0.025. The numerical parameters used are $L=240$ lead states, $\tau_c=3.2$ with $N_s=4$ and $\delta t=0.8$. 
Note that convergence and thus agreement with Monte Carlo cannot be achieved for 
$t \geq 10$ if $\frac{U}{\Gamma}  \geq 3$.}
\label{Fig1}
\end{figure}

\begin{figure}[htbp]
\vspace{0mm} \hspace{0mm}
 {\hbox{\epsfxsize=90mm \epsffile{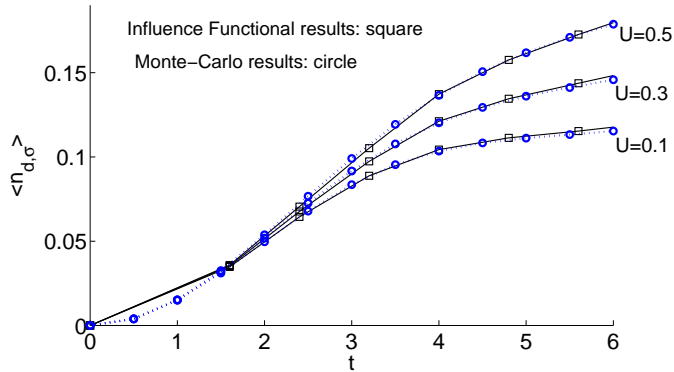}}}
 \caption{
Short time dynamics in the  Anderson model ($\square$) compared with
Monte Carlo data ($\circ$) for $U=0.5,0.3,0.1$ (top to bottom).
$D=1$, $\Delta \mu=0.4$, $E_d=0.3$,  $\Gamma_{\alpha}$=0.025,
$L$=240, $\tau_c=3.2$ with $N_s=4$ and $\delta t=0.8$. }
 \label{Fig1a}
\end{figure}

Figure \ref{Fig1} presents the time evolution of $\langle
n_{d,\sigma}\rangle$ with increasing on-site interaction.  While we
have not been able to overcome convergence issues for all times and
all values of $\frac{U}{\Gamma}$, we find that dynamics are
faithfully reproduced for all $\frac{U}{\Gamma}$ at short times,
while accurate and converged results are correctly obtainable only for
$\frac{U}{\Gamma}\lesssim  3$. 
The strict requirements for convergence are presented in Appendix D.
While this regime is one where perturbation theory in $U$ is accurate \cite{Phillip2,Lothar}, 
we believe that convergence restrictions are surmountable 
within the methodology presented in this work.  Future study will be devoted to this issue.
Fig. \ref{Fig1a} compares the early
propagation obtained within the IF approach ($\square$) to the MC
data ($\circ$). Interestingly, while our approach does not capture
the $t^2$ characteristic at $0<t<3$ due to the rough time
discretization, the intermediate time dynamics is still correct. It
should be possible to devise an adaptive time propagation scheme
where the time step is increasing with time, keeping $\tau_c$ fixed.
Future work will be devoted to improving convergence for large $U$
and $t$. It is interesting to note that even though the results at
large time and on-site energy ($U/\Gamma\gtrsim 3$)
 are not converged and thus do not
controllably represent a reliable estimate of population dynamics,
the results are still reasonably close to the MC data even for
$\frac{U}{\Gamma}=6$.


\section{Conclusions}

We have presented here a general path-integral based iterative
scheme for studying the dissipative dynamics of bias-driven
nonequilibrium systems. Our method relies on the finite range of
bath correlations in out-of-equilibrium cases, thus interactions
within the influence functional may be truncated beyond a memory time
dictated by the nonequilibrium conditions, and an iterative and deterministic scheme may be developed.  This scheme is in principle exact for cases where convergence with respect to truncation of memory effects is achieved.

The philosophy of our approach is similar to the previously developed ISPI approach of Thorwart, 
Egger and coworkers \cite{Egger}. 
The distinction between the method presented here and ISPI is confined to the propagation scheme 
and the technique via which the leads are eliminated.  
The discretized BSS-like approach \cite{BSS} to tracing out the reservoirs used here may be employed in 
situations where the structure of the memory term is difficult to obtain analytically. 
Furthermore, the matrices involved in the iterative scheme are fixed in size, and this fact may present 
numerical advantages at very long times. While our approach introduces an additional source of 
systematic error related to discretizing the leads, we have found that this error is easily controlled 
with limited numerical cost. Thus, our approach presents a related but complimentary methodology to the 
ISPI technique. It should be noted that currently the approach presented here and the ISPI technique appear 
to have difficulty converging in similar regions of parameter space that are accessible in some cases by, 
for example, the weak-coupling Monte Carlo approach \cite{Phillip2}. 
However approaches like ISPI and the methodology presented allow for an accurate description of 
long-time dynamical features when they do converge, something that is generically difficult with Monte Carlo schemes. 
In this regard our approach is also complimentary to, and not competitive with, 
expansion based Monte Carlo schemes \cite{Phillip,Phillip2}.

We have applied our technique to two prototype models: (i)
The spin-fermion model of a spin coupled via a
dipole-type interaction to two leads under a potential bias, and
(ii) the Anderson model, where a resonant level with an onsite
repulsion is coupled to nonequilibrium leads. In the first case
the dynamics of the tunneling system was investigated, recovering
damped oscillations for weak-intermediate couplings with the bias
playing a role analogous to that of the temperature in equilibrium
systems. For the nonequilibrium Anderson model we focused our study
on the resonant level population. Our method yields results in
reasonable agreement with numerically exact Monte Carlo simulations
for weak to intermediate onsite interactions $U$. For strong $U$
deviations are observed. 
The results presented in Appendix D suggest that the  deviations are related to memory and time step 
truncation errors which we have been unable to control at the present time. 
Future work will be devoted to this issue. 
The study of more complex models, e.g. the multilevel Anderson model 
with onsite electron-phonon interactions will be the subject of future studies.

\acknowledgments DS acknowledges support from the Connaught grant.
AJM was supported by NSF under Grant No.~DMR-0705847.
DRR would like to acknowledge the NSF for financial support.
The authors acknowledge P. Werner for fruitful discussions
and for providing the Monte Carlo data and
M. Thorwart for useful correspondence and encouragement.

\renewcommand{\theequation}{A\arabic{equation}}
\setcounter{equation}{0}  
\section*{Appendix A: Justification of the Truncation Scheme}

Here, we justify the breakup of the IF as prescribed by Eq.
(\ref{eq:IFbreak}), demonstrating that the terms neglected account
for interactions beyond the memory range $\tau_c$. Consider for
simplicity the functional
\bea
I(s_0^{\pm},s_1^{\pm},s_2^{\pm},s_3^{\pm},s_4^{\pm},s_5^{\pm})\approx
 I(s_0^{\pm},s_1^{\pm},s_2^{\pm},s_{3}^{\pm}) \frac
{I(s_1^{\pm},s_2^{\pm},s_3^{\pm}, s_{4}^{\pm}) }
{I(s_1^{\pm},s_2^{\pm},s_3^{\pm})} \frac
{I(s_2^{\pm},s_3^{\pm},s_4^{\pm}, s_{5}^{\pm}) }
{I(s_2^{\pm},s_3^{\pm},s_4^{\pm})}, \label{eq:IApp} \eea
truncated here by following Eq. (\ref{eq:IFbreak}) with $N_s$=3.
Using a cumulant expansion for the total influence functional (IF)
\cite{Multiple,Makri-An}, we write the IF as a product of $n$-body
interaction terms, $I=I(2) \times  I(3) \times I(4) \times I(5)$,
where each term is an exponent of a sum of the $n$-body terms, 
For example, $I(2)\sim e^{-\sum_{i,j}g_{i,j} }$ with pairwise
interactions $g_{i,j}$, $I(3)\sim e^{-\sum_{i,j,k} g_{i,j,k}}$,
incorporating "three body" interactions $g_{i,j,k}$. Substituting this
structure into Eq. (\ref{eq:IApp}), we find that the following terms
are not present on the right hand side: The two- and three-body
terms $g_{0,4}$, $g_{0,1,4}$, $g_{0,2,4}$, $g_{0,3,4}$, four-body
terms, $g_{0,1,2,4}$, $g_{0,1,3,4}$ and $g_{0,2,3,4}$, and a
five-body element $g_{0,1,2,3,4}$. These nonlocal interactions,
connecting spins beyond the memory range specified, $N_s=3$, are
assumed to be small, and are therefore discarded in our truncation
scheme.  
Larger memory blocks, connecting more distant time slices, may systematically be 
included until convergence with truncation of memory terms is reached.

To make this discussion concrete, consider a situation where
non-equilibrium Coulomb gas behavior holds, as discussed in
\cite{Marcus,MitraC}.  In such cases, the total influence functional
will be of the form $I\sim \exp\left[ \sum_{i>j}
C_0(|t_i-t_j|)\right]$ where $C_0(t) \propto \Delta \mu |t|$ up to
logarithmic corrections.  Consider now Eq. (\ref{eq:IFbreak}).  Clearly the leading term
contains all interactions between ``charges'' separated by a distance
in time that does not exceed $|t_{0}-t_{N_s}|$, namely
$I(s_0^{\pm},s_1^{\pm},...,s_{N_s}^{\pm}) \sim \exp\left[
\sum^{N_{s}}_{i>j}\sum^{N_s-1}_{j=0} C_0(|t_i-t_j|)\right]$. Terms of
the form
$\frac{I(s_1^{\pm},s_2^{\pm},...,s_{N_s+1}^{\pm})}{I(s_1^{\pm},s_2^{\pm},...,s_{N_s}^{\pm})}$
include only interactions between ``charges'' interacting over the
time intervals $|t_{n}-t_{N_s+1}|$ where $0<n< N_{s+1}$, without
double counting terms already contained in
$I(s_0^{\pm},s_2^{\pm},...,s_{N_s}^{\pm})$.  This procedure is then
iteratively continued until the complete influence functional is
constructed.  The error accrued originates from the neglect of 
terms in the exponent of the order $\Delta \mu \tau$ where $\tau
=|t_{a}-t_{b}|$ and $b-a \geq N_s+1$.  Thus, the procedure is rendered
controlled and is expected to converge to the exact result as long as
$N_s$ is made large enough.  It should be noted that the approach
outlined here is more general than this and is expected to hold at
short times or very large couplings where Coulomb gas behavior may
break down, as discussed in \cite{Marcus,MitraC}.

\renewcommand{\theequation}{B\arabic{equation}}
\setcounter{equation}{0}  
\section*{Appendix B: Extensions of the IF Technique to Finite Temperatures}

We present here the natural extension of our approach to finite temperature. The core of our numerical calculation is the influence functional (IF), incorporating the Fermi sea degrees of freedom, e.g. Eq. (\ref{eq:IFSF}) for the spin-fermion model or
Eq. (\ref{eq:IFAM}) for the Anderson model.  Assuming for simplicity a single Fermi sea, consider the following IF-like object
\bea
C_f={\rm Tr}_B \left[e ^{M_1} e^{M_2} \rho_B \right],
\eea
where $M_1$ and $M_2$ are quadratic operators and $\rho_B=e^{-\beta H_B}/{\rm Tr}_B[e^{-\beta H_B}]$,
$H_B$ is the bath Hamiltonian, (\ref{eq:HB}).
This correlation function can be expressed by single-particle operators \cite{Levitov},
\bea
C_f={\rm det} \left[ I-f(\epsilon) +e^{m_1}e^{m_2} f(\epsilon)\right].
\eea
Here $f(\epsilon)=[1+e^{-\beta (\epsilon-\mu)}]^{-1}$
is the Fermi-Dirac distribution function, $\beta$ is the inverse temperature, $I$ is the unit operator,  and $m_1$ and $m_2$ are single-particle operators corresponding to $M_1$ and $M_2$ respectively.
This expression can be trivially extended to include more exponential terms, $e ^{M_1} e^{M_2}... \cdot e ^{M_N}$, as necessary for the evaluation of the IF expression. For multiple-independent reservoirs, $\rho_B=\rho_L \otimes \rho_R$, the above relation can be generalized,
\bea
C_f&=&{\rm Tr}_L {\rm Tr}_R \left[e ^{M_1} e^{M_2} \rho_L\otimes  \rho_R \right]
\nonumber\\
&=& {\rm det} \left\{ \left[ (I_L-f_L(\epsilon)) \otimes I_R\right]  \left[ (I-f_R(\epsilon))\otimes I_L  \right]
+ e^{m_1}e^{m_2} \left[ f_L(\epsilon) \otimes I_R \right] \left[ f_R(\epsilon)\otimes I_L\right] \right\}.
\nonumber\\
\eea
Here $I_{\alpha}$ is the identity matrix for the $\alpha$ space;  $\alpha=L,R$,
and $f_\alpha(\epsilon)=[1+e^{-\beta_{\alpha} (\epsilon-\mu_{\alpha})}]^{-1}$.  The above expressions reduce to the ones used in the text for $T=0$.

\renewcommand{\theequation}{C\arabic{equation}}
\setcounter{equation}{0}  
\section*{Appendix C: An Alternative Formulation: Nonequilibrium Transport Through an Anderson Dot}

We present here an alternative formulation for calculating the dot properties in the single impurity Anderson model (SIAM) without invoking the Hubbard-Stratonovich transformation. This formulation is based on a different Trotter decomposition than that used in Section IV.  While the resulting expressions are more complex for the decomposition described here, it has the advantage that the resulting IF need not be updated each time step. Furthermore, since fewer terms of the Hamiltonian are split in the Trotter decomposition, it is possible that larger time steps may be taken with the decomposition presented here.  Further work investigating this approach, which is not confined to the Anderson model, will be presented in a future work.
We refer to the approach developed in Section IV as SIAM I, and
to the method of this appendix as SIAM II.

We begin by partitioning the Hamiltonian (\ref{eq:SIAM}) as follows:
$H_0$ includes the subsystem (dot) terms, and $H_1$ includes
the two non-interacting leads ($H_B$) and system-bath couplings ($H_{SB}$)
\bea
H^{AM}&=&H_0+H_1, \,\,\,\,\,\  H_1=H_B+ H_{SB}
\nonumber\\
H_{0}&=&\sum_{\sigma} \epsilon_d n_{d,\sigma}  +
U n_{d,\uparrow} n_{d,\downarrow},
\nonumber\\
H_B&=&\sum_{\alpha,k,\sigma}\epsilon_k
c_{\alpha,k,\sigma}^{\dagger}c_{\alpha,k,\sigma}; \,\,\, H_{SB}= \sum_{\alpha,k,\sigma}V_{\alpha,k}c_{\alpha,k,\sigma}^{\dagger}d_{\sigma}
+h.c.
\label{eq:ASIAM}
\eea
Here $n_{d,\sigma}=d^{\dagger}_{\sigma}d_{\sigma}$ is the impurity number operator and $c_{\alpha,k,\sigma}^{\dagger}$ is
a creation operator of an electron   at the $\alpha$ lead  with a spin $\sigma$ and momentum $k$.
Note that $H_0$ can be explicitly described
by a 4-state system, $|1\rangle=|0,0\rangle$, $|2\rangle=|\uparrow,0\rangle$, $|3\rangle=|\downarrow,0\rangle$, $|4\rangle=|\uparrow,\downarrow\rangle$,
corresponding to an empty dot, a single occupied dot of $\sigma=\uparrow, \downarrow$, and a double occupancy state.
When $U$ is very large ($U\rightarrow \infty$), we effectively have a 3-state system, since double occupancy becomes negligible. The energies of these four subsystem states are $E_1=0$, $E_{2,3}=\epsilon_d$, and $E_{4}=\epsilon_d+U$.

Consider the reduced density matrix $\rho_S={\rm Tr_B}\{\rho\}$ obtained by tracing the total density matrix
$\rho$ over the reservoir degrees of freedom.
The time evolution of $\rho_S(t)$ is exactly given by
\bea
\rho_S(a,a',t)={\rm Tr_B}\langle a| e^{-iH^{AM}t} \rho(0) e^{iH^{AM}t} |a'\rangle,
\label{eq:ArhoS}
\eea
where $|a\rangle$ and $|a'\rangle$ are subsystem states, as described above.
Using the standard Trotter breakup, $e^{iHt}=\left(  e^{iH\delta t}\right)^N$, $\delta t=t/N$, and
$e^{iH^{AM}\delta t} \approx e^{iH_0\delta t/2} e^{iH_1\delta t}  e^{iH_0\delta t/2}$,
we can rewrite Eq. (\ref{eq:ArhoS}) in a path integral formulation,
\bea
&&\rho_S(a,a',t)=
\int ds_0^+ \int ds_1^+ ... \int ds_{N-1}^+\int ds_0^- \int ds_1^- ... \int ds_{N-1}^-
\nonumber\\
&&{\rm Tr_B}
\Big\{\langle a| e^{-iH_0\delta t/2} e^{-iH_1\delta t} e^{-iH_0\delta t/2}|s^+_{N-1}\rangle
\langle s_{N-1}^+| e^{-iH_0\delta t/2} e^{-iH_1\delta t} e^{-iH_0\delta t/2}|s^+_{N-2}\rangle ...
\nonumber\\
&& \langle s_0^+| \rho(0)|s_0^{-}\rangle ...
\langle s^-_{N-2}| e^{iH_0\delta t/2} e^{iH_1\delta t} e^{iH_0\delta t/2}|s^-_{N-1}\rangle
\langle s_{N-1}^-| e^{iH_0\delta t/2} e^{iH_1\delta t} e^{iH_0\delta t/2}|a'\rangle
\Big \},
\label{eq:AEOM}
\eea
where $s_k$ are subsystem states.
As an initial condition we may assume that
$\rho(0)=\rho_{B}\rho_S(0)$ with the bath ($B$) uncoupled to the subsystem. We focus next on the following matrix elements in Eq. (\ref{eq:AEOM})
\bea
G_{a,b}(\delta t)\equiv \langle a|  e^{-iH_0\delta t/2} e^{-iH_1\delta t} e^{-iH_0\delta t/2} |b \rangle
=e^{-i(E_a+E_b)\delta t/2} \langle a|  e^{-iH_1\delta t}   |b \rangle.
\label{eq:AG}
\eea
%
To compute $ \langle a|  e^{-iH_1\delta t}   |b \rangle$ note that it is advantageous to use again the Trotter splitting
\bea
\langle a|  e^{-iH_1\delta t}   |b \rangle \approx
e^{-iH_{B}\delta t/2}  \langle a|  e^{-iH_{SB}\delta t}   |b \rangle   e^{-iH_{B}\delta t/2}.
\label{eq:AV}
\eea
We thus focus next on the matrix element
\bea
\hat O_{a,b}=\langle a|e^{-iH_{SB} \delta t}|b \rangle,
\label{eq:AO}
\eea
a quadratic operator in the space of
the non-interacting electrons. It is useful to define the "composite" fermion $c_{0,\sigma}=\sum_{\alpha,k} V_{\alpha,k} c_{\alpha,k,\sigma}$,
leading to $H_{SB} \equiv \sum_{\sigma} \left( c_{0,\sigma}^{\dagger} d_{\sigma} + d_{\sigma}^{\dagger} c_{0,\sigma}\right)$.
In this representation a direct expansion of the exponential gives
\bea
e^{\lambda H_{SB}} = I +(\cosh \lambda-1)\hat a_2 +\sinh\lambda \hat a_1
\label{eq:AGE}
\eea
with $\lambda=-i\delta t$,  $\hat a_1 =H_{SB}$ and $\hat a_2=\sum_{\sigma} \left( d_{\sigma }d_{\sigma}^{\dagger}
c_{0,\sigma}^{\dagger} c_{0,\sigma} d_{\sigma } d_{\sigma}^{\dagger} + h.c. \right)$.
The operator (\ref{eq:AO})  is therefore of the form, $\hat O_{a,b} = \alpha +\beta c_{0,\sigma} + \beta'c_{0,\sigma}^{\dagger}
+\gamma c_{0,\sigma}^{\dagger}c_{0,\sigma}+ \gamma' c_{0,\sigma}c_{0,\sigma}^{\dagger}$,
with constant coefficients $\alpha,\beta,\gamma$.
Substituting the pieces (\ref{eq:AV})- (\ref{eq:AGE}) into Eq. (\ref{eq:AG}) yields
\bea
G_{a,b}(\delta t)\approx e^{-i(E_a+E_b)\delta t/2} e^{-iH_{B}\delta t/2} \hat O_{a,b}  e^{-iH_{B}\delta t/2},
\eea
incorporating linear combinations of bath operators $c_{0,\sigma}$ up to a quadratic order.
Finally, we put all pieces together into Eq. (\ref{eq:AEOM}) and obtain the reduced dynamics

\bea
&&\rho_S(a,a',t)=
\int ds_0^+ ... \int ds_{N-1}^+\int ds_0^-  ... \int ds_{N-1}^-
\langle s_0^+ |\rho_S(0) |s_0^-\rangle
\nonumber\\
&&\exp{\big[ -i\delta t\sum_{j=1}^{N-1} E_{s_j^+}    +i\delta t\sum_{j=1}^{N-1}E_{s_{j}^-}
-i(E_a+E_{s_0^+})\delta t/2+ i(E_a'+E_{s_0^-})\delta t/2  \big]}
\nonumber\\
&&\times
{\rm Tr_B} \Big \{ \left( e^{-iH_{B}\delta t/2} \hat O_{a,s_{N-1}^+}  e^{-iH_{B}\delta t}
 \hat O_{s_{N-1}^+,s_{N-2}^+} e^{-iH_{B}\delta t} ...
 \hat O_{s_1^+,s_0^+} e^{-iH_{B}\delta t/2}  \right)
 \rho_B(0)
 \nonumber\\
&& \left(    e^{-iH_{B}\delta t/2} \hat O_{s_{0}^-,s_1^-} e^{-iH_{B}\delta t}
  \hat O_{s_1^-,s_2^-} e^{-iH_{B}\delta t} ...
  \hat O_{s_{N-1}^-,a'} e^{-iH_{B}\delta t/2}
  \right)
\Big \}.
\label{eq:AEOM2}
\eea
Identifying the integrand as the IF, we can
use the approach of Section II, define the truncated IF $I_s$, and
iteratively propagate the reduced density matrix to long times.

The approach developed here (SIAM II) has three main advantages over the method described in the main text (SIAM I),
see  Section IV. First, since the present method does not rely on the Hubbard-Stratonovich transformation it can be applied to general
many body interaction Hamiltonians, while  SIAM I is restricted to the Anderson model. Second, since the Trotter error in SIAM II is due to system-bath factorization, rather than one-body- many-body
splitting as in SIAM I, the method described here should be beneficial
in calculating dynamics of weakly coupled system-bath models with arbitrarily large many body (local) interactions.
Finally, this method also suggests a computational advantage over SIAM I, since the IF here [integrand of Eq. (\ref{eq:AEOM2})]
is  {\it time independent}, unlike the IF of Eq. (\ref{eq:IFAM}) which needs to be recalculated at each time step.

\renewcommand{\theequation}{D\arabic{equation}}
\setcounter{equation}{0}  
\section*{Appendix D:  Convergence Analysis for the Anderson Model}

There are three separate sources of systematic error within our
approach. (i)  {\it Bath discretization error}. The electronic
reservoirs are explicitly included in our simulations, and we use
bands extending from $-D$ to $D$ with a finite number of states per
bath per spin ($L$). This is in contrast to standard approaches
where a wide-band limit is assumed and analytical expressions for
the reservoirs Green's functions are adopted
 \cite{Rabani,Egger,Phillip}. (ii) {\it Trotter error}. The time
discretization error originates from the approximate factorization
of the total Hamiltonian into the non-commuting $H_0$ (two-body) and
$H_1$ (many-body) terms, see text after Eq. (\ref{eq:At}). While for
$U\rightarrow 0$ and for small time-steps $\delta t\rightarrow 0$
the decomposition is exactly satisfied, for large $U$ one should go
to a sufficiently small time-step in order to avoid significant
error buildup. (iii) {\it Memory error}. Our approach assumes that
bath correlations exponentially decay resulting from the
nonequilibrium condition  $\Delta \mu \neq 0$. Based on this crucial
element, the influence functional may be truncated to include only
a finite number of fictitious spins $N_s$, where $\tau_c=N_s \delta
t \sim 1/\Delta \mu$. The total IF is retrieved by taking the limit
$N_s \rightarrow N$, ($N=t/\delta t$).

These three errors can be systematically eliminated by increasing
the number of bath states, choosing a small enough time-step, and
adopting a sufficiently long memory time. Note however that the last
two strategies are linked: Increasing $\tau_c$ essentially means
increasing the time-step, since the memory length is restricted to
small values $N_s=4-6$ for practical-computational reasons. Thus, as
in standard QUAPI \cite{QUAPI}, one should find an optimal balance
between the time-step error and the memory size that correctly
represents the dynamics. Ref. \cite{Thorwart} suggests a systematic
approach for reaching convergence using the QUAPI method,
eliminating the Trotter discretization error and the memory
truncation inaccuracy by extrapolating the data to vanishing
time-step and to infinite memory time.

\begin{figure}[htbp]
\vspace{0mm} \hspace{0mm}
{\hbox{\epsfxsize=90mm \epsffile{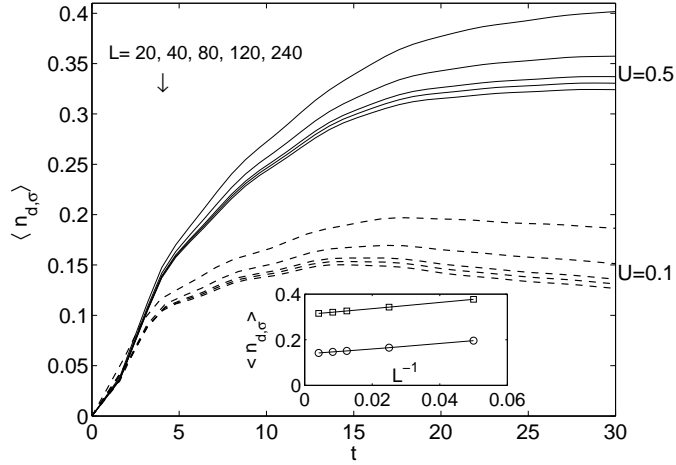}}}
\caption{Convergence of the dot occupancy with increasing number of bath states $L$.
$E_d=0.3$, $\Gamma_{\alpha}$=0.025, $N_s=4$, $\tau_c=3.2$.
Full lines (top to bottom); $U=0.5$,  $L$ =20, 40, 80, 120, 240;
Dashed lines (top to bottom):
$U=0.1$,  $L$ =20, 40, 80, 120, 240.
Inset: data as a function of $L^{-1}$ at $t=20$,
$U=0.5$ (square); $U=0.1$ (circle).}
\label{Fig4}
\end{figure}
\begin{figure}[htbp]
\vspace{0mm} \hspace{0mm} {\hbox{\epsfxsize=90mm
\epsffile{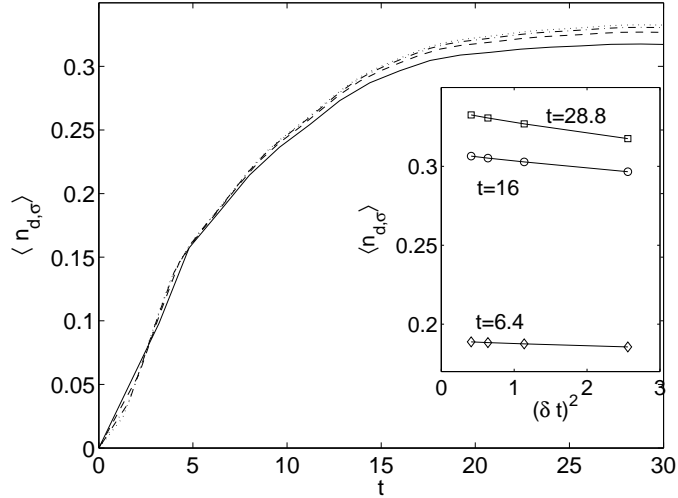}}} \caption{Convergence of dot occupancy reducing
the time-step $\delta t=\tau_c/N_s$. $E_d=0.3$, $U=0.5$,
$\Gamma_{\alpha}$=0.025,  $\tau_c=3.2$, $L$=120, $\delta t=1.6$
(full); $\delta t=1.07$ (dashed); $\delta t=0.8$ (dashed-dotted);
$\delta t=0.64$ (dotted). Inset: Data as a function of $(\delta
t)^2$ for three representative times.} \label{Fig5}
\end{figure}
\begin{figure}[htbp]
\vspace{0mm} \hspace{0mm} {\hbox{\epsfxsize=95mm
\epsffile{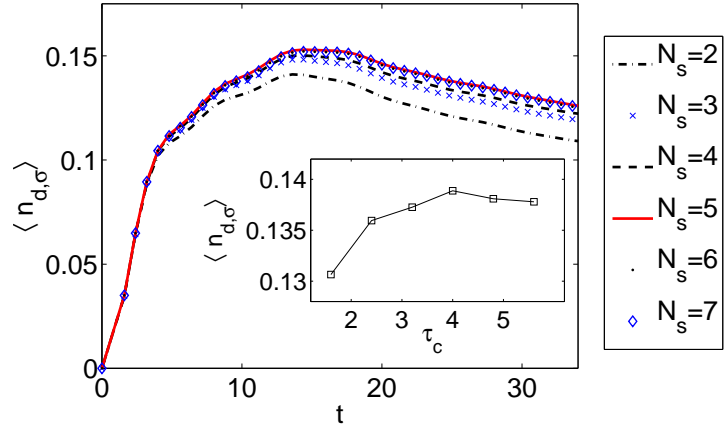}}} \caption{Convergence of dot occupancy with
increasing memory size $\tau_c$. 
$E_d=0.3$, $U=0.1$,
$\Gamma_{\alpha}$=0.025, $L$=120, $\delta t=0.8$.
$N_s$=2 (dashed-dotted); $N_s$=3 ($\times$); $N_s$=4 (dashed);
$N_s$=5 (full); $N_s$=6 (o);  $N_s$=7 ($+$)
Inset: Dot population vs. $\tau_c$ at a specific time, $t=12$.}
\label{Fig6}
\end{figure}

A similar idea can be adopted here. First the bath finite-size error
can be eliminated by systematically increasing the number of
fermions at each lead. As an example, Fig. \ref{Fig4} presents the
dot population for $U=0.1$ and $U=0.5$ taking $L$= 20, 40, 80, 120
and 240 (top to bottom). The inset shows that convergence can be
reached, and that the occupancy is systematically decreasing with
$L$. Next, the Trotter error can be eliminated by extrapolating the
data to the $\delta t \rightarrow 0$ limit. Fig. \ref{Fig5} presents
as an example the occupancy for $\Delta\mu=0.4$ using $\tau_c=3.2$,
and $\delta t=1.6, 1.05, 0.8, 0.64$. The inset manifests convergence
as a function of $(\delta t)^2$. Note that in the asymptotic limit
the data points are slightly enhanced, practically canceling the
effect of the bath discretization. Finally, the memory effect is
analyzed in Fig. \ref{Fig6}. For the parameters employed here
($E_d=0.3$, $U=0.1$, $\Gamma_{\alpha}=0.025$, $\Delta\mu=0.4$)
convergence is arrived at $\tau_c\sim 4$ (inset), in agreement with the
rough estimate $\tau_c\sim 1/\Delta \mu$. We have not been able to obtain full convergence for $U/\Gamma \geq 3$.

Using this analysis, we have recalculated  Fig. \ref{Fig1}
extrapolating our data to (i) $L\rightarrow \infty$, (ii) $\delta
t\rightarrow 0$ and (iii) $\tau_c\rightarrow \infty$. Since the
extrapolations (i) and (ii), bring about counter contributions, see
Figs. \ref{Fig4} and \ref{Fig5}, the overall effect of the bath-time
step-memory extrapolations on the occupation is rather small, and
Fig. \ref{Fig1} remains essentially intact.

\begin{thebibliography}{32}
\bibitem{MATLAB}
Simulations were performed here using the MATLAB compiler and
the Parallel Computing Toolbox of MATLAB.
\end{thebibliography}
\end{document}